\documentstyle[twocolumn,prb,aps,floats,epsfig]{revtex}
\begin{document}
\twocolumn[\hsize\textwidth\columnwidth\hsize\csname 
@twocolumnfalse\endcsname
\title{Piezoresistive anisotropy of percolative granular metals} 
\author{C. Grimaldi$^1$, P. Ryser$^1$, and S. Str\"assler$^{1,2}$} 
\address{$^1$ Institut de Production et Robotique, LPM,
Ecole Polytechnique F\'ed\'erale de Lausanne,
CH-1015 Lausanne, Switzerland}
\address{$^2$ Sensile Technologies SA, PSE, CH-1015 Lausanne, Switzerland}

\maketitle

\centerline \\

\begin{abstract}
The piezoresistive response of granular metals under uniaxial strain
is strongly dependent on the concentration of the conducting phase.
Here we show that the piezoresistive anisotropy is  reduced
as the system approaches its percolation thresold, following a power
law behavior in the critical region. We propose a simple relation between the
conductance and the piezoresistive anisotropy which could be used
in relation to real materials and notably to the thick film resistors.
PACS numbers: 72.20.Fr, 72.60.+g, 72.80.Ng
\end{abstract}
\vskip 2pc ] 

%\narrowtext
\centerline \\

Disordered systems constituted by random dispersions of metallic and 
insulating phases, also commonly referred to as granular metals, show transport
properties strongly dependent of the concentration $x$ of the metallic phase in
the composite. For $x$ larger than a critical concentration $x_c$, transport
is metallic-like while for $x \rightarrow x_c$ the system undergoes a metal-to-insulator
transition. In the critical region $(x-x_c)\ll 1$, the conductance $G$
follows a power-law behavior $G\sim (x-x_c)^t$ where $t$ is the critical
exponent. The minimal model of a theoretical descriptions of transport in
granular metals is based on extensions of percolation theory to random-resistors 
network models.\cite{kirk} 
Numerical simulations of such percolative resistors show that as
$x$ approaches $x_c$ from above, fluctuations of the microscopic (bond)
currents get enhanced. It has been shown recently that this phenomenon
can be spoiled to get more accurate experimental determinations of $x_c$
compared to common conductance measurements.\cite{chan}

The enhancement of bond current fluctuations is basically also at the
origin of the peculiar behavior of the piezoresistance effect as 
$x\rightarrow x_c$.\cite{grima1} Under an uniaxial imposed strain, in fact, the
piezoresistive response is highly anisotropic when $x\simeq 1$, 
being maximum for external electric fields applied along the direction of 
the imposed strain (longitudinal response) and minimum along the
orthogonal directions (transverse response). This is due to the fact that
for $x\simeq 1$ the microscopic bond currents are basically all in the
direction of the external electric field.
Instead, lower values of $x$ force the bond currents to have components
also along directions perpendicular to the external fields so that
the anisotropy of the piezoresistive response is gradually
reduced until at $x=x_c$ the longitudinal and transverse responses
become equal.\cite{grima1} Hence, at the percolation thresold the piezoresistive 
effect is perfectly isotropic.

The transport properties of granular metals under uniaxial strain
could in principle help to characterize the vicinity to the percolation thresold
with an higher accuracy compared to that typical of common
conductance measurements. The interest on the piezoresistive response
of granular metals however concerns also more applicative problems.
In fact, a particular class of granular metals made of RuO$_2$,
IrO$_2$ or other metal-oxide granules embedded in a glassy matrix,
also commonly known as thick-film resistors (TFRs),
are characterized by a quite large piezoresistive response, which is
used in fabricating highly sensitive pressure and force sensors.\cite{white,prude1}
It is thought that intergrain tunneling processes are responsible
for the TFRs high strain sensibility.\cite{pike,prude2}
A characteristic common to most of the TFRs is that their piezoresistive
response is quasi-isotropic,\cite{prude1,prude2,hrovat} indicating that the values and
directions of the microscopic currents are highly fluctuating functions
and that the TFRs are quite close to the percolation thresold.
This is confirmed also by the observation of power-law dependence of
$G$ upon RuO$_2$ concentrations.\cite{kusy}

In this communication, we compute the piezoresistive response of different 
random resistor network models as a function of bond or site concentrations.
We show that close to the percolation thresold, the piezoresistive anisotropy 
follows an universal behavior governed by an exponential law.
Our prediction could be easily experimentally tested, and could bring
important informations on the topology of the percolative network and
the critical point of TFRs.

To model the piezoresistive behavior of TFRs, or more generally of 
percolative granular metals, we consider a three-dimensional 
cubic random-resistor network
whose bond conductances either zero or proportional to a tunneling
exponential factor $\exp(-2d_i/\xi)$, where $d_i$ is the tunneling distance
between two neighbouring sites along the $i$ direction 
and $\xi$ the localization length. 
In our model, 
the network bond directions are parallel to the orthogonal $x$, $y$, $z$ axes,
and in the absence of applied strains the tunneling distances are all equal:
$d_i=d$, $i=x,y,z$. In this situation, the network is isotropic and the
total conductance $G_i=G$ depends only on the specific bond conductance distribution.
To study the effect of an applied strain, let us assume that the resistor
network is embedded in an homogeneous elastic medium and that the elastic 
coefficients of the network and the medium are equal. Hence, when a strain field
$\varepsilon_{ii}$ ($i=x,y,z$) is applied to the sample,
the conductance changes to $G_i=G+\delta G_i$, where the variation 
$\delta G_i$ can be expressed
in terms of the (intrinsic) conductivity variation $\delta \sigma_i$ and a geometric
(extrinsic) factor:\cite{prude3}
\begin{equation}
\label{G1}
\frac{\delta G_i}{G}=\frac{\delta \sigma_i}{\sigma}-\varepsilon_{ii}+
\varepsilon_{jj}+\varepsilon_{kk},
\end{equation}
where $\sigma$ is the unstrained conductivity and the indexes $i$, $j$, $k$ 
assume the values $x$, $y$, and $z$ with cyclic permutations.
In the above equation, the variation $\delta\sigma_i$ is due to
the tunneling distance change $d\rightarrow d_i=d(1+\varepsilon_{ii})$ which 
leads to
$\exp(2d/\xi)\rightarrow \exp(2d/\xi)(1+2d\varepsilon_{ii}/\xi)$ for a
conducting bond along the $i$ direction.

\begin{figure}[t]
\protect
\centerline{\epsfig{figure=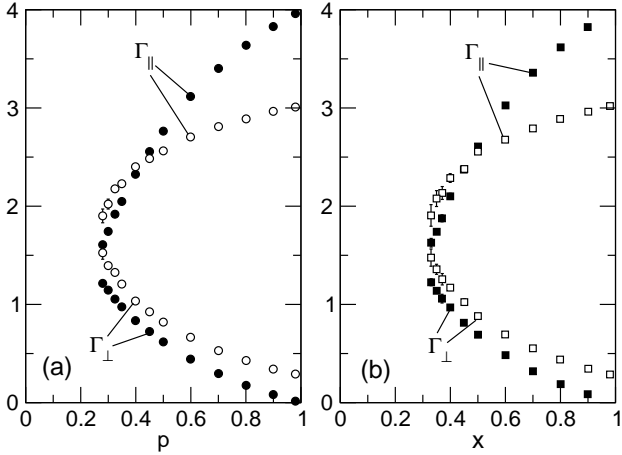,width=20pc,clip=}}
\caption{Monte Carlo calculations of longitudinal
($\Gamma_\parallel$) and transverse ($\Gamma_\perp$) piezoresistive
coefficients of bond percolation (a) and site percolation (b) cubic network models.
The calculations have been performed for networks of $L^3$
sites with $L=30$ and $L=40$ by numerically solving iteratively 
the Kirchhoff node equations (for more details see Ref.\protect\onlinecite{grima1}). 
Filled symbols refer to bond
conductances with tunneling exponent $2d/\xi=4$, while open
symbols are the results of uniform random distribution
$2\leq 2d/\xi \leq 6$. In (a), $\Gamma_\parallel$ and $\Gamma_\perp$
become equal at the critical bond concentration $p=p_c\simeq 0.248$.
In (b), piezoresistive isotropy is reached at the critical site
concentration $x=x_c\simeq 0.314$ (which corresponds to $p_c=x_c^2\simeq 0.098$).}
\label{fig1}
\end{figure}

To express $\delta \sigma_i/\sigma$ as a function of a general $\varepsilon_{ii}$, let
us first imagine that an uniaxial strain along the $x$ 
direction ($\varepsilon_{xx}=\varepsilon$, $\varepsilon_{yy}=\varepsilon_{zz}=0$) is applied
to the sample. In this situation, the conductivity along the $x$ axis will in general be different
from those along the $y$ and the $z$ axes which, by symmetry, are instead equal.
Therefore, up to linear terms in $\varepsilon$,
$\sigma_x=\sigma-\sigma\Gamma_\parallel\varepsilon$ and
$\sigma_y=\sigma_z=\sigma-\sigma\Gamma_\perp\varepsilon$ where
\begin{eqnarray}
\label{gammapar}
\Gamma_\parallel &=&-\frac{\delta \sigma_x}{\varepsilon \sigma}, \\
\label{gammaper}
\Gamma_\perp &=&-\frac{\delta \sigma_y}{\varepsilon \sigma}=
-\frac{\delta \sigma_z}{\varepsilon \sigma},
\end{eqnarray}
are the longitudinal and transverse piezoresistive coefficients, respectively.
The above reasoning can be repeated for uniaxial strains along the $y$ and the $z$ axis and,
since the problem is linear, for a general strain field $\varepsilon_{ii}$ the 
conductivity variations $\delta \sigma_i$ reduce to:
\begin{equation}
\label{G2}
\frac{\delta \sigma_i}{\sigma}=-\Gamma_\parallel\varepsilon_{ii}-
\Gamma_\perp(\varepsilon_{jj}+\varepsilon_{kk}). 
\end{equation}
The above equation permits to express all the piezoresistive coefficients
$\Gamma_{ij}=-\delta\sigma_i/\varepsilon_{jj}\sigma$ or, by using Eq.(\ref{G1}),
the corresponding piezoresistive gauge
factors $K_{ij}=-\delta G_i/\varepsilon_{jj}G$ (that are the commonly measured quantities)
in terms of $\Gamma_\parallel$ and $\Gamma_\perp$. 
For example, in a typical experiment with a thick cantilever beam having the main axis 
directed along the $x$ direction, the strains are approximatively 
$\varepsilon_{xx}=\varepsilon$, $\varepsilon_{yy}=-\nu\varepsilon$,
and $\varepsilon_{zz}=-\nu'\varepsilon$, where $\nu$ and $\nu'$ are the Poisson ratios
of the cantilever and the resistive sample, respectively.\cite{prude3}
By using Eqs.(\ref{G1},\ref{G2}), the longitudinal ($K_{\rm L}$) and transverse (K$_{\rm T}$)
piezoresistive gauge factors are:
\begin{eqnarray}
K_{\rm L}\equiv K_{xx}&=&(1+\Gamma_\parallel)+(1-\Gamma_\perp)(\nu+\nu'), \\
K_{\rm T}\equiv K_{yx}&=&-(1+\Gamma_\parallel)\nu 
-(1-\Gamma_\perp)(1-\nu') .
\end{eqnarray}
Hence from a measurement of $K_{xx}$ and $K_{yx}$ it is possible to extract
the {\it intrinsic} piezoresistive coefficients $\Gamma_\parallel$ and $\Gamma_\perp$.
As we are going to show below, the knowledge of $\Gamma_\parallel$ and $\Gamma_\perp$ 
permits to characterize the nature of transport of percolative piezoresistors.

As already pointed out in the introduction, the piezoresistive response depends in a non trivial way
on the vicinity of the resistor network to its percolation critical point. 
Consider in fact Fig. 1a in which we report numerical Monte Carlo calculations of $\Gamma_\parallel$
and $\Gamma_\perp$ for bond percolating random resistor networks. In this model, a fraction
$p$ of bonds distributed at random has conductance $\exp(-2d/\xi)$, while the remaining fraction
$1-p$ has zero conductance. In the figure, filled symbols refer to a model in which
the tunneling exponent $2d/\xi$ is set equal to $4$, while the open circles are obtained
by treating $\xi$ as a random variable in such a way that $2 \leq 2d/\xi\leq 6$.\cite{grima1}
When there are no missing bonds ($p=1$) the piezoresistive response
is anisotropic with the longitudinal coefficient much larger than the transverse one,
reflecting the fact that the current flows mainly along paths parallel to the direction
of the voltage drop. This is more evident for the $2d/\xi=4$ data (filled circles) for
which $\Gamma_\parallel=2d/\xi=4$ and $\Gamma_\perp=0$ at $p=1$. If fluctuations in
the values of $2d/\xi$ are allowed (empty circles) even at $p=1$ the microscopic currents
develop a component ortogonal to the voltage drop direction 
leading to $\Gamma_\parallel < 2d/\xi$ and $\Gamma_\perp>0$.
When bonds are removed ($p<1$), the longitudinal response gets reduced and at the same time
the transverse piezoresistive coefficient is enhanced. This trend persists all the way
down to the percolation thresold ($p=p_c\simeq 0.2488$) where
$\Gamma_\parallel$ equals $\Gamma_\perp$ and the system shows perfect piezoresistive
isotropy. As shown in Fig.1b, the same qualitative behavior holds true also for site 
percolation models in which a concentration $x$ of sites is removed at random with
remaining conducting bonds having
$2d/\xi=4$ (filled squares) or $2 \leq 2d/\xi\leq 6$ (open squares). 

From the results of both bond percolation and site percolation models
of Fig.1 we can infer that as a general rule $\Gamma_\parallel\rightarrow\Gamma_\perp$ when the resistor 
network approaches its percolation thresold. This behavior is better analyzed in Fig. 2 where
we plot the piezoresistive anisotropy factor 
$(\Gamma_\parallel-\Gamma_\perp)/\Gamma_\parallel$ obtained from the data of Fig.1 as a 
function of $p-p_c$. For both bond and site percolation models with or without distribution of
$2d/\xi$ values, the anisotropy factor monotonically goes to zero as $p\rightarrow p_c$
and in the critical region $p-p_c\ll 1$ it follows a power law behavior:
\begin{equation}
\label{power1}
\frac{\Gamma_\parallel-\Gamma_\perp}{\Gamma_\parallel}\sim (p-p_c)^\lambda.
\end{equation}
This specific power law dependence is made more clear in the inset of Fig. 2 where
the data are plotted in a log-log scale. A fit of the numerical data to Eq.(\ref{power1}) 
leads to $\lambda=0.7\pm 0.1$ for bond percolation and $\lambda=0.4\pm 0.2$ for site percolation
independently of $2d/\xi$ within the error bars. Although we obtain a critical exponent for
bond correlation somewhat larger than that for site percolation, we do not take this difference too
seriously since a precise determination of $\lambda$ requires much more extensive numerical
calculations than those presented here. This is outside the scope of this communication, which
instead hints to point out the existence of a critical exponent $\lambda$ for the
piezoresistive anisotropy.

\begin{figure}[t]
\protect
\centerline{\epsfig{figure=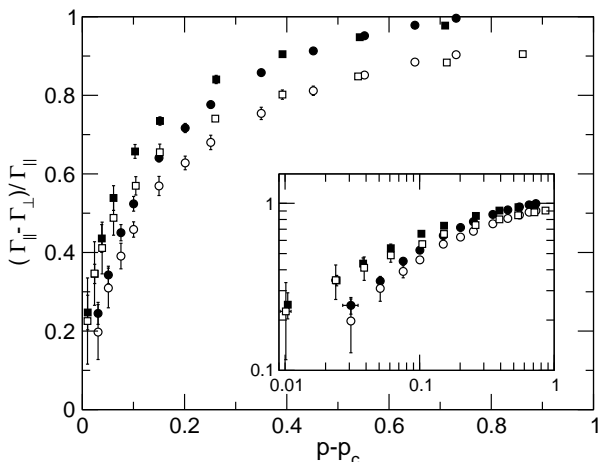,width=20pc,clip=}}
\caption{Piezoresistive anisotropy factor as a function of $p-p_c$
for the bond (circles) and site (squares) percolation data of Fig. 1. In the inset the
same data of the main figure are plotted in a log-log scale.}
\label{fig2}
\end{figure}

The power law behavior of Eq.(\ref{power1}) can also be inferred from the transport
properties of anisotropic bond percolation models studied some
time ago.\cite{shklo,lobb,sary} In those works, percolating networks with random
bond conductances were defined in such a way that the conducting bonds were equal 
along two directions, for example $y$ and $z$, but different from those along the 
third direction, that is $x$. Topological considerations,\cite{shklo}
renormalization group analysis,\cite{lobb} and numerical calculations\cite{sary}
showed that the quantity $\sigma_y/\sigma_x-1=\alpha(p)$ has a power law behavior 
$\alpha(p)\sim(p-p_c)^{\lambda'}$ close to the percolation thresold. By neglecting 
the geometrical factors, it is easy to see from Eqs.(\ref{gammapar},\ref{gammaper}) 
that in our piezoresistive
model $\sigma_y/\sigma_x-1\simeq (\Gamma_\parallel-\Gamma_\perp)\varepsilon$, so that
the two exponents $\lambda$ and $\lambda'$ are equal.

In summary, we have shown numerically that in the critical regime of percolative resistor networks
the piezoresistive anisotropy has a power law behavior. This result suggests
a possible experimental route
to characterize transport properties in TFRs or other piezoresistive granular
compounds. In fact, as pointed out in the introduction, typical TFRs show
a power law behavior of the conductance $G$ as a function of metallic concentration.\cite{kusy}
Hence, we predict that if this behavior is due to the closeness to a percolative
critical point, then the piezoresistive anisotropy factor should follow Eq.(\ref{power1})
or equivalently:
\begin{equation}
\label{power2}
G\sim\left(\frac{\Gamma_\parallel-\Gamma_\perp}{\Gamma_\parallel}\right)^{\lambda/t},
\end{equation}
where $t$ is the critical exponent of the unstrained conductance $G\sim (p-p_c)^t$. We are
not aware of published data from which it is possible to infer $G$ and
$(\Gamma_\parallel-\Gamma_\perp)/\Gamma_\parallel$ as a function of the metallic 
concentration in a controlled way, so Eq.(\ref{power2}) cannot yet be tested. 
However, there is evidence that the piezoresistive gauge factors 
anisotropy $(K_{\rm L}-K_{\rm T})/K_{\rm L}$ in some
commercial TFRs is lower for higher resistive samples,\cite{hrovat} in qualitative accord
therefore with Eq.(\ref{power2}). We encourage acquisition of additional experimental data
in TFRs to verify our predictions.


\begin{references}

\bibitem{kirk}
S. Kirkpatrick, Rev. Mod. Phys. {\bf 45}, 574 (1973).

\bibitem{chan}
M.-C. Chan, A. B. Pakhomov, and Z.-Q. Zhang,
J. Appl. Phys. {\bf 87}, 1584 (2000).

\bibitem{grima1}
C. Grimaldi, P. Ryser, and S. Str\"assler, 
Phys. Rev. B {\bf 64}, 064201 (2001).

\bibitem{white}
N. M. White and J. D. Turner,
Meas. Sci. Technol. {\bf 8}, 1 (1997).

\bibitem{prude1}
M. Prudenziati, {\it Handbook of Sensors and Actuators}
(Elsevier, Amsterdam, 1994), p.189.

\bibitem{pike}
G. E. Pike and C. H. Seager,
J. Appl. Phys. {\bf 48}, 5152 (1977).

\bibitem{prude2}
C. Canali, D. Malavasi, B. Morten,
M. Prudenziati, and A. Taroni, J. Appl. Phys. {\bf 51},
3282 (1980).

\bibitem{hrovat}
M. Hrovat, J. Holc, D. Belavi\v{c}, and S. \v{S}oba,
J. Mater. Sci. Lett. {\bf 13}, 992 (1994)

\bibitem{kusy}
A. Kusy, Physica B {\bf 240}, 226 (1997).

\bibitem{prude3}
B. Morten, L. Pirozzi, M. Prudenziati, and A. Taroni,
J. Phys. D: Appl. Phys. {\bf 12}, L51 (1979).

\bibitem{shklo}
B. I. Shklovskii, Phys. Status Solidi (b) {\bf 85}, K111 (1978).

\bibitem{lobb}
C. J. Lobb, D. J. Frank, and M. Tinkham,
Phys. Rev. B {\bf 23}, 2262 (1981).

\bibitem{sary}
A. K. Sarychev and A. P. Vinogradoff,
J. Phys. C: Solid State Phys. {\bf 12}, L681 (1979);
{\it ibid.} {\bf 16}, L1073 (1983).

\end{references}
\end{document}